\documentclass[reqno,11pt]{amsart}

\usepackage{amsmath,amssymb}

\newtheorem{theorem}{Theorem}
\newtheorem{lemma}{Lemma}
\newtheorem{corollary}{Corollary}

\theoremstyle{remark}
\newtheorem{remark}{Remark}

\newenvironment{acknowledgement}{\par\medskip\noindent\emph{Acknowledgement.}}

\DeclareMathOperator{\tr}{tr}

\let\HH\H
\renewcommand\H{\mathcal{H}}

\newcommand\R{\mathbb R}

\newcommand\Z{\mathbb Z}

\renewcommand\d{\mathrm d}
\newcommand\e{\mathrm{e}}
\newcommand\eps{\varepsilon}

\newcommand{\la}{\langle}
\newcommand{\ra}{\rangle}

\renewcommand\P{\mathbb P}
\newcommand\E{\mathbb E}

\newcommand{\scal}[1]{\la #1 \ra}

\def\Chi{\raisebox{.4ex}{$\chi$}}
\DeclareMathOperator{\spann}{span}

\DeclareMathOperator{\dom}{dom}
\DeclareMathOperator{\supp}{supp}
\DeclareMathOperator{\dist}{dist}

\def\le{\leqslant}
\def\ge{\geqslant}

\begin{document}

\title[Persistence of Anderson localization]
{Persistence of Anderson localization in Schr\"odinger operators with decaying
  random potentials}

\author[A. Figotin]{Alexander Figotin}
\address{University of California, Irvine,
  Department of Mathematics,
  Irvine, CA 92697-3875,  USA}
\email{afigotin@uci.edu}
 
\author[F. Germinet]{Fran\c cois Germinet}
\address{ Universit\'e de Cergy-Pontoise,
  D\'epartement de Math\'ematiques,
  Site de Saint-Martin,
  2 avenue Adolphe Chauvin,
  95302 Cergy-Pontoise cedex, France}
\email{germinet@math.u-cergy.fr}

\author[A. Klein]{Abel Klein}
\address{University of California, Irvine,
  Department of Mathematics,
  Irvine, CA 92697-3875,  USA}
\email{aklein@uci.edu}
 
\author[P. M\"uller]{Peter M\"uller}
\address{Institut f\"ur Theoretische Physik,
Georg-August-Universit\"at G\"ottingen,
Friedrich-Hund-Platz 1,
37077 G\"ottingen, Germany}
\email{peter.mueller@physik.uni-goe.de}

\thanks{A.\ Figotin was sponsored by the Air Force Office of Scientific 
Research Grant FA9550-04-1-0359.}
\thanks{A.\ Klein was  supported in part by NSF Grant DMS-0457474.}
\thanks{P.\ M\"uller was supported in part by
  the Deutsche Forschungsgemeinschaft under Grant Mu~1056/2--1.}

%\thanks{Version of \today}

\begin{abstract}
  We show persistence of both Anderson and dynamical localization in
  Schr\"odinger operators with non-positive (attractive) random
  decaying potential.  We consider an Anderson-type Schr\"odinger
  operator with a non-positive ergodic random potential, and multiply
  the random potential by a decaying envelope function.  If the envelope function decays slower than $|x|^{-2}$ at
  infinity,  we 
  prove that the operator has  infinitely many eigenvalues below
  zero. For envelopes decaying as $|x|^{-\alpha}$ at infinity, we
  determine the number of bound states below a given energy $E<0$,
  asymptotically as $\alpha\downarrow 0$. To show that bound states
  located at the bottom of the spectrum are related to the phenomenon
  of Anderson localization in the corresponding ergodic
  model, we prove: ~(a)~ these states are exponentially localized with
  a localization length that is uniform in the decay exponent
  $\alpha$; (b)~ dynamical localization holds uniformly in $\alpha$.
\end{abstract}

\maketitle
\thispagestyle{empty}

%%%%%%%%%%%%%%%%%%%%%%%%%%%%%%%%%%%%%%%%%%%%%%%%%%%%%%%%%%%%
%%%%%%%%%%%%%%%%%%%%%%%%%%%%%%%%%%%%%%%%%%%%%%%%%%%%%%%%%%%%

\section{Introduction and results}

A mathematical proof of the existence of absolutely continuous (or
just continuous) spectrum for a multidimensional Schr\"odinger
operator with random ergodic potential is still a challenge.  Up to
date there is no proof of \emph{any} continuous spectrum for ergodic
random Schr\"odinger operators in $d$-dimensional spaces, neither on the
lattice nor in the continuum.  The only known result is the existence
of absolutely continuous spectrum for the Anderson model on the Bethe
lattice \cite{Kle98} (see also \cite{ AiSi05, FrHa05}).  The only
proof of existence of a localization-delocalization transition in
finite dimensions for a typical ergodic random Schr\"odinger operator
is for random Landau Hamiltonians ($d=2$), where non-trivial transport
has been shown to occur near each Landau level \cite{GKS}.  (See
\cite{JSBS} for a special delocalization phenomenon in one-dimensional
random polymer models.)

To gain insight into this fundamental question, one may impose a
decaying envelope on the ergodic random potential, and study the
absolutely continuous spectrum for the new Schr\"odinger operator with
random decaying potential as a step towards the understanding the
original problem \cite{Kr,KKO,Bo1,Bo2,RoSc,De,BSS,Che05}.  Relaxing
 the decay conditions, one hopes to get an idea of the
nature of the continuous spectrum for the original ergodic random
Schr\"odinger operator. If the imposed envelope decays fast enough,
regular scattering theory applies, and one may conclude that the
spectrum is absolutely continuous regardless of the randomness. This
indicates that the essence of the original problem is to establish the
existence of continuous spectrum ``in spite of the randomness'' of the
ergodic potential.  Since randomness leads to Anderson localization
and the existence of non-trivial pure point spectrum, one must answer
the question of when continuous spectrum can coexist with Anderson
localization. In particular, we may ask if this coexistence phenomenon
can already be seen in Schr\"odinger operators with random decaying
potential.
 
In this paper we show persistence of both Anderson and dynamical
localization in Schr\"odinger operators with non-positive (attractive)
random decaying potential.  We consider the random Schr\"odinger
operator
\begin{equation}
  \label{defH}
  H_{\alpha,\lambda,\omega} := -\Delta + \lambda \gamma_{\alpha}
  V_\omega \quad \text{on} \quad  
  \mathrm{L}^{2}(\mathbb{R}^{d}),
\end{equation}
where $\lambda > 0$ is the disorder parameter, $\alpha \ge 0$,
$\gamma_{\alpha}$ is the \emph{envelope function}
\begin{equation}
  \label{alpot}
  \gamma_{\alpha}(x) :=  \langle x\rangle^{-\alpha}, \quad
  \text{where}\quad   \langle x\rangle:=\sqrt{1+ |x|^{2}} ,
\end{equation}
and $V_\omega$ is the \emph{non-positive} random potential given by
\begin{equation}\label{defV}
  V_\omega(x) := - \sum_{j\in\mathbb{Z}^{d}} \omega_{j} \, u(x-j).
\end{equation}
Here $\{\omega_{j}\}_{j\in\mathbb{Z}^{d}}$ are independent identically
distributed random variables on some probability space $(\Omega,
\mathbb{P})$, with $0 \le \omega_{0} \le 1$ and
$\mathbb{E}\{\omega_0\} >0$.  The single-site potential $ u \in
\mathrm{L}^{\infty}(\mathbb{R}^{d})$ is assumed to satisfy
\begin{equation}\label{defu}
  0 \le u \le u_0 , \qquad 
  \supp u \mbox{ compact}, \qquad
  v:= \int_{\mathbb{R}^{d}} \!\d x\; u(x) >0 ,
\end{equation}
with $u_0>0$ a constant. We note that the support of $u$ may be
arbitrarily small. Under these hypotheses $ H_{\alpha,\lambda,\omega}$
is self-adjoint on the domain of the Laplacian $\Delta$ for every
$\omega\in\Omega$.

In the special case of a constant envelope function, obtained by
setting $\alpha =0$ in \eqref{alpot}, $
H_{\lambda,\omega}:=H_{0,\lambda,\omega}$ is the usual Anderson-type
Schr\"odinger operator with a non-positive ergodic random potential.
Due to ergodicity, the spectrum $\sigma(H_{\lambda,\omega})$ of
$H_{\lambda,\omega}$, as well as the spectral components in the
Lebesgue decomposition, do not depend on $\omega$ for
$\mathbb{P}$-almost every $\omega\in\Omega$ \cite{CaLa90, PaFi92}.  If
the single-site distribution $\mathbb{P}(\omega_{0} \in \cdot)$ has a
bounded Lebesgue density, it is also well-known that
$H_{\lambda,\omega}$ exhibits Anderson localization, both spectral and
dynamical, in a neighborhood above the non-random bottom
$E_{0}(\lambda) <0$ of its spectrum \cite{CH,GK3,St}.  The latter is
also true if $\omega_{0}$ is a Bernoulli random variable
($\mathbb{P}(\omega_{0} =0)=\mathbb{P}(\omega_{0} =1)=\frac 12 $), and
may be shown by modifying \cite{BK} as in \cite{GHK2}.
%
%The problem of proving existence of absolutely continuous spectrum
%gets easier if the random potential is very weak at infinity.  For
%this reason, Schr\"odinger operators with \emph{non-ergodic} random
%potentials have drawn a lot of attention during the past years. Sparse
%random potentials and potentials with decaying randomness are of
%particular importance among them
%\cite{Kr,KKO,Bo1,Bo2,RoSc,De,BSS,Che05}.  
For non-ergodic random Schr\"odinger operators, like
$H_{\alpha,\lambda,\omega}$ with $\alpha>0$, one cannot expect
non-randomness of the spectrum and of the spectral components in
general.

If $\alpha >1$, one is able to construct wave operators for
$H_{\alpha,\lambda,\omega}$. This implies that for all $\lambda >0$
the absolutely continuous spectrum of $H_{\alpha,\lambda,\omega}$
coincides with $[0,\infty[$ for $\mathbb{P}$-almost all
$\omega\in\Omega$ \cite{HuKi} (see also \cite{Kr}). Recent results
suggest that this should be true for all $\alpha > 1/2$ \cite{Bo1,
  Bo2,De}.  Of course, the primary interest in models like
\eqref{defH} is for small parameters $\alpha$, when they are ``close''
to the ergodic random Schr\"odinger operator $H_{\lambda,\omega}$.

Since the random potential in \eqref{defV} is non-positive,
$H_{\alpha,\lambda,\omega}$ can only have discrete spectrum at
energies below zero: its essential spectrum is almost surely equal to
$[0,\infty[$, cf. \cite[Thm.\ II.4.3]{CaLa90} or \cite[Ex.~6 in Ch.\
XIII.4]{RS}. Consequently, for any given $\alpha>0$ the random
operator $H_{\alpha,\lambda,\omega}$ exhibits localization of
eigenfunctions and even dynamical localization in any given interval
below zero for $\mathbb{P}$-almost every $\omega\in\Omega$. But does
this localization regime have anything in common with the well-studied
region of complete localization---as it was called in
\cite{GK4,GKS}---that occurs for $\alpha=0$, i.e., for the
corresponding ergodic random Schr\"odinger operator
$H_{\lambda,\omega}$?

To get insight into this question, suppose a bound state of
$H_{\alpha,\lambda,\omega}$ with energy $E<0$ was localized solely
because of the presence of the envelope. Then it would be localized in
a ball of size $|E|^{-1/\alpha}$, roughly.  Outside this ball, in the
classically forbidden region, it would decay exponentially fast.
Hence, the slower the decay of the envelope, the weaker this type of
localization would be.  In particular, it would disappear in the limit
$\alpha\downarrow 0$. Our main result, given in part~(3) of
Theorem~\ref{main}, shows that this is not the case. Localization
occurs uniformly in $\alpha$ so that bound states of
$H_{\alpha,\lambda,\omega}$ are localized because of the presence of
randomness, and not because of the decaying envelope. Likewise,
dynamical localization holds uniformly in $\alpha \ge 0$. This is not
a trivial property either, because we show in part~(2) of
Theorem~\ref{main} that the number of contributing eigenfunctions
diverges as $\alpha \downarrow 0$. Thus, Anderson localization
persists also from a dynamical point of view.

In the formulation of Theorem~\ref{main}, we use the notation
\begin{equation} 
  n(A,E) :=  \#\bigl\{E_j \le E:
  E_j \mbox{ is an eigenvalue of } A \bigr\}
\end{equation} 
for the number of eigenvalues of a self-adjoint operator $A$ which do
not exceed a given $E\in\mathbb{R}$---counted according to their
multiplicities. (This number is always finite  if $A$ is bounded from
below and has only discrete spectrum up to $E$.) 

\begin{theorem} \label{main} 
 Let $H_{\alpha,\lambda,\omega}$ be as in \eqref{defH}--\eqref{defu}.
  \leftmargini1.6em
  \leftmarginii1.6em
  \begin{enumerate}
  \item 
    If $\alpha \in]0,2[$, then $H_{\alpha,\lambda,\omega}$ has
    infinitely many eigenvalues in $]-\infty, 0[$ for
    $\mathbb{P}$-almost every $\omega\in\Omega$.
  \item 
    Let $E_{0}(\lambda) <0$ denote the non-random bottom of the
    spectrum of $H_{\lambda,\omega}$.  For $\mathbb{P}$-almost every
    $\omega\in\Omega$, the inequalities
    \begin{align}
      \hspace*{.7cm}d \log \tfrac{1}{\nu_{0}(\lambda,E)}
      \le \liminf_{\alpha\downarrow 0} \; \bigl[ \alpha \log
      n(H_{\alpha,\lambda,\omega},E) \bigr]  
      & \le\limsup_{\alpha\downarrow 0} \; \bigl[ \alpha \log
      n(H_{\alpha,\lambda,\omega},E) \bigr] \nonumber\\  
      & \le  d \log \left(\tfrac{\lambda U_{0}}{|E|}\right)
      \label{nlimit}
    \end{align}
    hold for all $E \in]E_{0}(\lambda),0[$, where $\nu_{0}(\lambda,E)
    := \inf\{\nu \in ]0,1[ : E_{0}(\nu\lambda) < E\}$ and $U_{0}:=
    \|\sum_{j\in\mathbb{Z}^{d}} u(\cdot -j)\|_{\infty}$.
  \item 
    If the single-site distribution $\mathbb{P}(\omega_{0} \in \cdot)$ 
    has a bounded Lebesgue density, then there exists an energy
    $E_{1}(\lambda) \in ]E_0(\lambda), 0[$ such that
    \begin{enumerate}
    \item 
      for $\mathbb{P}$-almost every $\omega\in\Omega$, any
      eigenfunction $\varphi_{n,\alpha,\lambda,\omega}$ of
      $H_{\alpha,\lambda,\omega}$ with eigenvalue in $ I_{\lambda} :=
      [E_{0}(\lambda), E_{1}(\lambda)]$ decays exponentially fast with
      a mass $m>0$.  The mass $m$ can be chosen independently of
      $\alpha\ge  0$.  More precisely, one has the following SULE-like
      property: there exists a localization center
      $x_{n,\alpha,\lambda,\omega}$ located in the ball centered at the origin
      and of radius $\mathcal{O}\bigl(|E|^{-\frac1\alpha}\bigr)$, if
      $|E|< 2\lambda u_0$ and $\alpha\le 1$, and
      $\mathcal{O}\bigl(|E|^{-1}\bigr)$ otherwise, such that for any
      $\varepsilon>0$,
      \begin{equation}\label{sule}
        \|\Chi_x \varphi_{n,\alpha,\lambda,\omega}\| \le
        C_{\varepsilon,\lambda,\omega} 
        \mathrm{e}^{|x_{n,\alpha,\lambda,\omega}|^\eps}
        \mathrm{e}^{-m|x-x_{n,\alpha,\lambda,\omega}|}
      \end{equation}
      for all $x\in\mathbb{R}^{d}$ and $\alpha \ge 0$, where
      $C_{\varepsilon,\lambda,\omega} >0$ is a constant independent of
      $\alpha$ and $\Chi_{x}$ is the indicator function of the unit
      cube in $\mathbb{R}^{d}$ centered at $x$;
    \item 
      one has uniform dynamical localization: for any $p\ge 0$,
      \begin{equation}\label{unifdynloc}
        \sup_{\alpha\ge 0} \; \sup_{|f| \le 1} \mathbb{E}
        \Bigl[\bigl\|\scal{x}^p 
        f(H_{\alpha,\lambda,\omega}) \Chi_{I_\lambda}(H_{\alpha,\lambda,\omega})
        \Chi_0 \bigr\|_2^2 \Bigr] < \infty ,
      \end{equation}
      where $\|\cdot\|_{2}$ stands for the Hilbert--Schmidt norm and the
      supremum is taken over all measurable functions $f: \mathbb{R}
      \rightarrow \mathbb{R}$ which are bounded by one.
    \end{enumerate}
  \end{enumerate}
\end{theorem}

\begin{remark}
  For $\alpha \in ]1,2[$ the operator $H_{\alpha,\lambda,\omega}$ has
  both absolutely continuous spectrum for energies in $[0,\infty[$
  (see the discussion above) and infinitely many eigenvalues below
  zero.
\end{remark}

\begin{remark}
  \label{refloc}
  It follows from the proof of part~(3) of the
  theorem in Section~\ref{sectdynloc} that the interval $I_\lambda$
  corresponds to the range of energies where one can prove
  localization for $H_{\lambda,\omega}$, that is, energies for which
  the initial-scale estimate of the multiscale analysis can be
  established for the corresponding ergodic operator. The rate of
  exponential decay also coincides with the one of the ergodic
  model. In other words, the eigenfunctions have the same localization
  length uniformly in $\alpha$.
\end{remark}

\begin{remark}
  In a few typical cases, one can show that the length of the interval
  $I_\lambda$ scales (at least) like $\mathcal{\lambda}$. At small
  disorder $\lambda$, this is proved in \cite{W,Kl2,Kl3}. At large
  disorder, this is shown in \cite{GK2,GK3} under the assumption that
  the single-site potential $u$ satisfies the covering condition $u
  \ge v_0 \Chi_{\Lambda_1} >0$ for some $v_{0} >0$.  Such an
  assumption can be removed using an averaging procedure as in
  \cite{BK,GHK}, in which case it is enough to assume that there
  exists $\delta>0$ and $v_0>0$ such that $u \ge v_0
  \Chi_{\Lambda_\delta}$, but the length of the interval $I_\lambda$
  then scales as $\lambda^{\rho}$ for some $\rho \in ]0,1[$.
\end{remark}

\begin{remark}
  We point out that our result does not apply to positive potentials,
  i.e., with a reversed sign in \eqref{defV}. For instance, the
  standard proof of the initial-scale estimate would fail for boxes
  that are far from the origin. In fact, at least for $\alpha$ large
  enough, one expects the absolutely continuous spectrum to fill up
  the entire positive half-line. The existence of a localized phase for
  low energies and small $\alpha$ is an open problem in this case, see
  \cite{Bo1}.
\end{remark}

\begin{remark}
  Dynamical localization is just one property of the region of
  complete localization, further properties can be found in
  \cite{GK2,GK4}. In particular, following \cite{GK4}, one can show
  decay of the kernel of the Fermi projector and strong uniform decay
  of eigenfunction correlations (SUDEC), uniformly in $\alpha$. We
  would like to emphasize that while for ergodic models these
  properties are known to be characterizations of the region of
  complete localization, i.e., they provide necessary and sufficient
  conditions, adding the envelope destroys the equivalence, and only
  the ``necessary part'' survives.
\end{remark}

This paper is organized as follows.
In Sections~\ref{sectinfev}, \ref{sectdensity}, and
\ref{sectdynloc}, we prove parts~(1), (2), and (3), respectively, of
Theorem~\ref{main}.

%%%%%%%%%%%%%%%%%%%%%%%%%%%%%%%%%%%%%%%%%%%%%%%%%%%%%%%%%%%%%%%%%%%%%%%%%%%%%%%
%%%%%%%%%%%%%%%%%%%%%%%%%%%%%%%%%%%%%%%%%%%%%%%%%%%%%%%%%%%%%%%%%%%%%%%%%%%%%%%
%%%%%%%%%%%%%%%%%%%%%%%%%%%%%%%%%%%%%%%%%%%%%%%%%%%%%%%%%%%%%%%%%%%%%%%%%%%%%%%

\section{Infinitely many bound states}
\label{sectinfev}

In this section we deduce part~(1) of Theorem~\ref{main} from a
corresponding result for slightly more general Schr\"odinger operators
with decaying random potentials. Given any non-negative function $ 0 \le
\gamma \in \mathrm{L}^{\infty}(\mathbb{R}^{d})$, we
consider  the random Schr\"odinger operator
\begin{equation}
  \label{defHg}
  H_{\omega}(\gamma) := -\Delta + \gamma V_\omega  \quad \text{on} \quad 
  \mathrm{L}^{2}(\mathbb{R}^{d}),
\end{equation}
 where  $V_{\omega}$ is  as in
\eqref{defV}. For \eqref{defHg} to represent a Schr\"odinger operator
with decaying randomness, we require that the envelope function
$\gamma$ vanishes at infinity, $\lim_{|x|\to\infty} \gamma(x)=0$.
Part~(1) of Theorem~\ref{main} then follows immediately from 

\begin{theorem}
  \label{thminfev}
  Let $H_\omega(\gamma)$ be as in \eqref{defHg}.  Suppose 
  \begin{equation}
    \label{gammaF}
    \gamma(x) |x|^{2} \ge F(|x|) \quad \text{for all $|x| > R_0$},
  \end{equation}
where   $R_0 >0$ and  $F\colon  [R_0,\infty[ \rightarrow ]0,\infty[$
  is a strictly increasing function such that 
  \begin{equation}
  \lim_{r\to\infty}F(r)
  = \infty \qquad  and  \qquad \lim_{r\to\infty}\frac {F(r) }{r^{2}} = 0.
  \end{equation} Then
  $H_\omega(\gamma)$ has infinitely many eigenvalues in $]-\infty, 0[$
  for $\mathbb{P}$-almost every $\omega\in\Omega$.
\end{theorem}

\begin{remark} 
  Theorem~\ref{thminfev} extends known deterministic results, namely
  \cite[Thm.~13.6]{RS} and \cite[Thm.~A.3(iii) and~(iv)]{DHS}, to the
  random case.  Thanks to randomness we do not have to require that
  each realization $V_{\omega}$ of the potential stays away from zero,
  whereas this has to be assumed for the deterministic potentials in
  \cite{RS, DHS}.  Note also \cite[Thm.~5.3(ii)]{DaHu03b}, which gives
  an infinite number of eigenvalues for discrete random Schr\"odinger
  operators on $\mathbb{Z}_{+}$ with an arbitrary (deterministic)
  potential subject to $\limsup_{n\to\infty} |V(n)| n^{\frac 12 } >2$.
\end{remark}

\begin{remark} 
  Theorem~\ref{thminfev} is almost optimal, because \cite[Thm.~A.3(i)
  and~(ii)]{DHS} (see also \cite[Thm.~13.6]{RS}) implies that
  $H_{\omega}(\gamma)$ has at most finitely many eigenvalues in
  $]-\infty, 0[$ for $\mathbb{P}$-almost every $\omega\in\Omega$,
  whenever $\gamma(x)\, |x|^{2} \le (1-\frac d 2)^{2}$ for all $|x|
  >R_{0}$ if $d\ge 3$, or whenever $\gamma(x)\, |x|^{2} \le
  (2\log|x|)^{-2}$ for all $|x| >R_{0}$ if $d=2$.
\end{remark}

\begin{proof}[Proof of Theorem~\ref{thminfev}]
  Without loss of generality, we shall assume that the support of $u$
  is included in $\Lambda_{1}$, where $\Lambda_{L}:=
  ]\frac L2,\frac L2[^d$ for $L>0$ (the smaller $\supp u$, the smaller
  is the number of eigenvalues). Also, on account of \eqref{gammaF},
  we may assume that $F$ grows to infinity as slowly as we want (the
  smaller $F$, the smaller the number of eigenvalues).  We shall prove
  that for all $L \ge L_{0} >2R_0$, with $L_{0}$ large
  enough and 
  depending on $d,v,R_{0}, \E(\omega_0)$, and $F$,
  there is a constant $c >0$ such that
  \begin{equation} 
    \mathbb{P} ( A_L)  
    \ge  1 - [F(L)]^{\frac d4} \e^{-c L^d [F(L)]^{-\frac d4}}
    ,\label{probbound}
  \end{equation}
  where
  \begin{equation}
    A_L := \Bigl\{ \omega\in\Omega : H_\omega(\gamma) \text{ has at least
      $ \kappa [F(L)]^{\frac d4}$ 
      eigenvalues in } ]-\infty,0[ \Bigr\} ,
  \end{equation}
 with $0< \kappa <1$  some constant depending only on $d$ and $R_{0}$.
  The theorem then follows by taking $F$ with a growth rate that is slow
  enough and using
  the Borel--Cantelli Lemma. 
  
  To prove \eqref{probbound},
  we set
  \begin{equation}\label{defell}
    \ell := [F(L)]^{-\frac14}L,
  \end{equation}
  and divide the shell $\Lambda_L\setminus \Lambda_{2R_0}$ into
  $N=\mathcal{O}[(L/\ell)^d - (2R_{0}/\ell)^{d}]$ non-overlapping
  cubes $\Lambda_\ell(n)$, $n=1,\ldots, N$, of side length $\ell$.
  (More precisely, we only consider the cubes contained in the shell.)
  Clearly, we have $ \kappa [F(L)]^{\frac d4} \le N \le [F(L)]^{\frac
    d4}$ with some $\kappa$ as above. For each $n=1, \ldots, N$ and
  each $\omega\in\Omega$ there exists a function $\varphi_{n} \in \dom
  (H_{\omega}(\gamma)) = \dom (-\Delta)$ such that
  \begin{enumerate}
  \item $\|\varphi_{n}\| =1$,
  \item $\varphi_{n}$ has compact support in $\Lambda_{\ell}(n)$,
  \item $\varphi_{n}|_{\Lambda_{\ell}^{\mathrm{int}}(n)} = c_{0}
    \ell^{-d/2}$, where $\Lambda_{\ell}^{\mathrm{int}}(n) := \{
    x\in\Lambda_{\ell}(n) : \dist_{\infty}(x,
    \partial\Lambda_{\ell}(n)) $ $\ge \ell/4\}$,
  \item $\|\,|\nabla\varphi_n|\,\|_{\infty} \le c_{0} \ell^{-1 -d/2}$,
  \end{enumerate}
  where the constant $c_{0} >0$ depends only on the dimension, and the
  distance in (3) is measured with respect to the maximum norm in
  $\mathbb{R}^{d}$. Note that the $\varphi_{n}$'s have disjoint
  supports, and hence are mutually orthogonal. From the above, and
  since $\Lambda_L \setminus \Lambda_{2R_0}$ is contained in the
  annulus with $|x| > R_{0}$ and $|x| \le \sqrt{d}L/2$, we conclude
  for every $n =1,\ldots,N$ and every $\omega\in\Omega$ that
  \begin{align} 
    \langle\varphi_{n}, H_{\omega}(\gamma) \varphi_{n}\rangle & \le
    \langle\varphi_{n}, \{ -\Delta + \widetilde{F}(L) L^{-2}
    V_{\omega}\} \varphi_{n}\rangle \nonumber \\
    & \le \| \nabla\varphi_{n}\|^{2} - \widetilde{F}(L) L^{-2} 
      \sum_{i\in\mathbb{Z}^{d}: \Lambda_1(i)\subset
        \Lambda_{\ell}^{\mathrm{int}}(n)} 
      \omega_{i} \langle \varphi_{n}, u(\cdot\, -i) \varphi_{n} \rangle
      \nonumber\\  
    & \le c_{0} \ell^{-2} - \widetilde{F}(L) L^{-2} 
      c_{0}^{2} \ell^{-d} v \sum_{i\in\mathbb{Z}^{d}: \Lambda_1(i)\subset
      \Lambda_{\ell}^{\mathrm{int}}(n)} 
      \omega_{i} ,
  \end{align}
  where $\widetilde{F}(L) := (4/d) F(\sqrt{d}L/2)$.  Recalling
  \eqref{defu} and the monotonicity of $F$, we infer the existence of
  a constant $c_{1}>0$ such that
  \begin{align}\label{bound2}
    \langle\varphi_{n}, H_{\omega}(\gamma) \varphi_{n}\rangle \le c_{0}
    \ell^{-2} - c_{1} \ell^{-2} [\widetilde{F}(L)]^{\frac12} v X_{n}^{(\omega)}
    (\ell),
  \end{align}
  where
  \begin{equation}\label{defX}
    X_{n}^{(\omega)} (\ell)
    := \frac{1}{Z_{n}(\ell)} \;       \sum_{i\in\mathbb{Z}^{d}:
      \Lambda_1(i)\subset \Lambda_{\ell}^{\mathrm{int}}(n)} \omega_{i} 
  \end{equation}
  and $Z_{n}(\ell)$ is the number of terms in the $i$-sum in
  \eqref{defX}.  Now, pick $0< \mu < \mathbb{E}\{\omega_{0}\}$. By a
  large-deviation estimate, there exists a constant $c_{2} >0$ such
  that
  \begin{equation} \label{ld}
    \mathbb{P} \Bigl\{ X_{n}^{(\omega)} (\ell) \ge \mu \text{~~for
      all~~} n=1,\ldots, N \Bigr\}  \ge 
    1 - \sum_{n=1}^{N} \, \mathbb{P} \bigl\{ X_{n}(\ell) \le \mu \bigr\} \ge
    1 - N \e^{-c_{2} \ell^{d}}
  \end{equation}
  holds for all $\ell$ sufficiently large. Thus, it follows from
  \eqref{bound2} and 
  \eqref{ld} that for $L$ large enough ensuring
  $c_{1} [\widetilde{F}(L)]^{\frac12} v\mu > c_{0}$, we have 
  \begin{equation}
    \label{maxprob}
    \mathbb{P}\biggl\{ \max_{n=1,\ldots,N} \langle\varphi_{n},
    H_{\omega}(\gamma) 
    \varphi_{n}\rangle < 0 \biggr\} \ge 1 - [F(L)]^{\frac d4}
    \e^{-c_{2} \ell^{d}}. 
  \end{equation}
  The bound \eqref{probbound} now follows from \eqref{maxprob} and the
  min-max principle. Indeed, we have the representation
  \begin{equation}
    \label{minmax}
    \lambda_{N}^{(\omega)} = \inf_{\mathcal{V}_{N} \subset
      \mathrm{L}^{2}(\mathbb{R}^{d}) \phantom{\|}} 
    \sup_{\psi\in\mathcal{V}_{N}: \|\psi\|=1} \langle\psi,
    H_{\omega}(\gamma)\psi\rangle \,,
  \end{equation}
  for the $N$th eigenvalue $\lambda_{N}^{(\omega)}$ (counted from the
  bottom of the spectrum and including multiplicities) of
  $H_{\omega}(\gamma)$. The infimum in \eqref{minmax} is taken over
  all $N$-dimensional subspaces $\mathcal{V}_{N}$ of Hilbert space.
  Therefore we have
  \begin{equation}
    \label{mmapply}
    \lambda_{N}^{(\omega)} \le \sup_{\psi\in \spann\{\varphi_{1},
      \ldots,\varphi_{N}\} :  
      \,\|\psi\|=1} \langle\psi, H_{\omega}(\gamma)\psi\rangle =
    \max_{n=1,\ldots,N} 
    \langle\varphi_{n}, H_{\omega}(\gamma)  \varphi_{n}\rangle <0,
  \end{equation}
  where we used the orthonormality of the $\varphi_{n}$'s, their
  disjoint supports, and the locality of $H_{\omega}(\gamma)$.
\end{proof}

%%%%%%%%%%%%%%%%%%%%%%%%%%%%%%%%%%%%%%%%%%%%%%%%%%%%%%%%%%%%%%%%%%%%%%%%%%%%%%%
%%%%%%%%%%%%%%%%%%%%%%%%%%%%%%%%%%%%%%%%%%%%%%%%%%%%%%%%%%%%%%%%%%%%%%%%%%%%%%%
%%%%%%%%%%%%%%%%%%%%%%%%%%%%%%%%%%%%%%%%%%%%%%%%%%%%%%%%%%%%%%%%%%%%%%%%%%%%%%%

\section{Counting the bound states}
\label{sectdensity}

In this section we deduce part~(2) of Theorem~\ref{main} from upper
and lower bounds on the number of eigenvalues of
$H_{\alpha,\lambda,\omega}$ in terms of the (self-averaging)
integrated density of states $N_{\lambda}$ of the ergodic random
Schr\"odinger operator $H_{\lambda,\omega}$.  

We recall \cite{CaLa90, PaFi92} that there exists a set
$\Omega_{0}\subset\Omega$ of full probability, $\mathbb{P}(\Omega_{0})
=1$, such that the macroscopic limit
\begin{equation}
  \label{dosdef}
  N_{\lambda}(E) := \lim_{L\to\infty} 
  \frac{n(H_{\lambda,\omega}^{(L,X)},E)}{L^{d}} 
\end{equation}
can be used to define a non-random, right-continuous and
non-decreasing function $N_{\lambda}$ on $\mathbb{R}$ in the sense
that \eqref{dosdef} holds for all $\omega \in\Omega_{0}$ and all
continuity points $E\in\mathbb{R}$ of $N_{\lambda}$. The operator
$H_{\lambda,\omega}^{(L,X)}$ in \eqref{dosdef} denotes the restriction
of $H_{\lambda,\omega}$ to the cube $\Lambda_L$. The boundary
condition $X$ can be arbitrary, as long as it renders the restricted
operator self-adjoint. In particular Dirichlet $(D)$ or Neumann $(N)$
boundary conditions are allowed and lead to the same integrated
density of states $N_\lambda$.

Since the envelope provides kind of an effective confinement for bound
states of $H_{\alpha,\lambda,\omega}$ with energy below zero, one
would expect
\begin{equation}
  \label{expect}
  \lim_{\alpha\downarrow 0} \frac{n(H_{\alpha,\lambda,\omega},
    E)}{l_{\alpha,\lambda, E}^{d}} = N_{\lambda}(E) 
\end{equation}
for all $E\in ]E_{0}(\lambda),0[$ and for $\mathbb{P}$-almost all
$\omega\in\Omega$, where $l_{\alpha,\lambda, E}$ is some effective,
non-random ``confinement length.''  We can prove the following.

\begin{theorem}
  \label{thmdensity}  Let $H_{\alpha,\lambda,\omega}$ be as in
  \eqref{defH} -- \eqref{defu}.
  Fix  $\lambda > 0$. For all $\delta>0$, $\nu\in]0,1[$,
  $\omega\in\Omega_{0}$ and $E <0$ there exists $\alpha_{0} > 0 $ such
  that for all $\alpha \in ]0,\alpha_{0}[$ we have
  \begin{equation}
    \label{nbounds}
    \left(\frac {\nu^{-\frac 2\alpha} -1}{d}\right)^{\frac{d}{2}} \;
    \bigl(N_{\nu\lambda}(E)-\delta \bigr) 
    \le
    \frac{n(H_{\alpha,\lambda,\omega},E)}{2^{d}}
    \le \left(\frac{\lambda U_{0}}{|E|}\right)^{\frac d\alpha}
    \bigl(N_\lambda(E)+\delta \bigr).
  \end{equation}
  The constant $U_{0}$ was defined in part~(2) of Theorem~\ref{main}.
  In particular, the estimates in  \eqref{nlimit} hold for all $E
  \in]E_{0}(\lambda),0[$ and all $\omega\in\Omega_{0}$.
\end{theorem}

\begin{corollary}
  \label{coincide}
  If, in addition, the single-site potential $u$ is chosen such that
  $E_{0}(\lambda) = -\lambda U_{0}$ (e.g.\ if $u$ is proportional to
  the characteristic function of the open unit cube around the origin,
  $u=U_{0} \Chi_{\Lambda_{1}}$), then we have $\nu_{0}(\lambda,E) =
  |E|/(\lambda U_{0})$, and \eqref{nlimit} implies the asymptotics
  \begin{equation}
    \lim_{\alpha\downarrow 0} \; \bigl[ \alpha \log
    n(H_{\alpha,\lambda,\omega},E) \bigr]   
    = d \log \left(\frac{\lambda U_{0}}{|E|}\right).
  \end{equation}
\end{corollary}

\begin{remark}
  If the limit \eqref{expect} exists in the situation of
  Corollary~\ref{coincide}, then the confinement length obeys
  $\lim_{\alpha\downarrow 0} \; \bigl[\alpha \log
  l_{\alpha,\lambda,E} \bigr] = \log (\lambda U_{0}/|E|)$.
\end{remark}

\begin{proof}[Proof of Theorem~\ref{thmdensity}]
  First, we turn to the lower bound in \eqref{nbounds}. Let $\alpha >0$,
  $\nu\in ]0,1[$ and set 
  \begin{equation}
    L_{\alpha}(\nu) := \frac{2}{\sqrt{d}} \; \left(\nu^{- \frac2\alpha}
    -1 \right)^{\frac{1}{2}}. 
  \end{equation}
  We observe that for every $\lambda>0$ and $\omega\in\Omega_{0}$, the
  (non-decaying) random potential of the Dirichlet restriction
  $H^{(L_{\alpha}(\nu), D)}_{0, \nu\lambda,\omega}$ is bounded from
  below by the decaying random potential of the Dirichlet restriction
  $H^{(L_{\alpha}(\nu), D)}_{\alpha,\lambda,\omega}$. This implies
  \begin{equation}
    n(H_{\alpha,\lambda,\omega},E)  \ge
    n(H_{\alpha,\lambda,\omega}^{(L_{\alpha}(\nu),D)},E)  
    \ge  n(H_{0,\nu\lambda ,\omega}^{(L_{\alpha}(\nu),D)},E)
  \end{equation}
  for all $E<0$. It follows from \eqref{dosdef} that for every $\delta>0$,
  there exists $\alpha_{0}^{-} >0$ (depending on $\delta, \nu, \lambda,
  \omega$ and $E$) such that,
  \begin{equation}\label{ne1}
    n(H_{\alpha,\lambda,\omega},E) \ge [L_\alpha(\nu)]^d
      [N_{\nu\lambda}(E) - \delta ]
  \end{equation}
  for all $\alpha \in ]0, \alpha_{0}^{-}[$.

  To obtain the upper bound in \eqref{nbounds}, we set
  $\ell_{\alpha}(E) := 2 |\lambda U_{0}/E|^{1/\alpha}$ for $\alpha
  >0$, $E<0$ and argue that the Neumann restriction of $H_{\alpha,
    \lambda, \omega }$ to $\mathbb{R}^{d} \setminus
  \Lambda_{\ell_{\alpha}(E)}$ cannot have any spectrum in $]-\infty,
  E]$ for every $\lambda >0$ and every $\omega\in\Omega_{0}$. This
  implies the first inequality in
  \begin{equation}
    \label{ne3}
    n(H_{\alpha,\lambda,\omega},E) \le
    n(H_{\alpha,\lambda,\omega}^{(\ell_{\alpha}(E), N)},E) 
    \le n(H_{0,\lambda,\omega}^{(\ell_{\alpha}(E), N)},E) ,
  \end{equation}
  the second one follows from monotonicity. Combining \eqref{ne3} and
  \eqref{dosdef}, we conclude that for every $\delta>0$ there exists
  $\alpha_{0}^{+} >0$ (depending on $\delta, \lambda, \omega$ and $E$) such
  that
  \begin{equation}\label{ne2}
    n(H_{\alpha,\lambda,\omega},E) 
    \le [\ell_{\alpha}(E)]^{d} \; [N_\lambda(E)+\delta]. 
  \end{equation}
  holds for all $\alpha \in ]0, \alpha_{0}^{+}[$.   Eqs.\ \eqref{ne1} and
  \eqref{ne2} prove \eqref{nbounds}. 

  As to the validity of \eqref{nlimit}, we remark that the condition
  $E_{0}(\nu\lambda) < E$, which enters through $\nu_{0}(\lambda,E)$,
  guarantees the positivity of the lower bound in \eqref{nbounds} for
  $\delta$ small enough.
\end{proof}

%%%%%%%%%%%%%%%%%%%%%%%%%%%%%%%%%%%%%%%%%%%%%%%%%%%%%%%%%%%%%%%%%%%%%%%%%%%%%%%
%%%%%%%%%%%%%%%%%%%%%%%%%%%%%%%%%%%%%%%%%%%%%%%%%%%%%%%%%%%%%%%%%%%%%%%%%%%%%%%
%%%%%%%%%%%%%%%%%%%%%%%%%%%%%%%%%%%%%%%%%%%%%%%%%%%%%%%%%%%%%%%%%%%%%%%%%%%%%%%

\section{Persistence of Anderson localization}
\label{sectdynloc}

In this section we prove part~(3) of Theorem~\ref{main} by a
multiscale argument. The fractional-moment method \cite{AENSS} should
work as well, provided the single-site potential $u$ satisfies the
covering condition $u \ge v_0 \Chi_{\Lambda_1} >0$.

%The multiscale analysis deals with ``finite volume operators'', which
%are restrictions of $H_{\alpha,\lambda,\omega}$ to finite volumes.
%These finite volume operators are required to have discrete spectrum
The multiscale analysis deals with restrictions of
$H_{\alpha,\lambda,\omega}$ to finite volumes. 
These ``finite volume operators'' are required to have discrete spectrum
in the range of energies we are interested in.  Since we work with
energies below the spectrum of the Laplacian, we choose
\begin{equation}\label{defHvol}
  H_{\alpha,\lambda,\omega}\bigl(\Lambda_L(x)\bigr) := -\Delta -
  \lambda\gamma_{\alpha} \sum_{i\in\Lambda_L(x)} \omega_i u(\cdot -i)
  =: -\Delta + V_{\Lambda_{L}(x)}
\end{equation}
(acting in $\mathrm{L}^{2}(\mathbb{R}^{d})$) for the restriction of the
operator $H_{\alpha,\lambda,\omega}$ to $\Lambda_L(x)$, the cube with
edges of length $L$ centered at $x\in\mathbb{R}^{d}$.

A crucial ingredient for the multiscale analysis is a Wegner
estimate. In the non-ergodic situation we are facing here, we need it
\emph{uniformly in the location of the center of the box}.
   
\begin{lemma}
  \label{lemwegner} \emph{(Wegner estimate)}\quad Assume that the
  single-site distribution $\mathbb{P}(\omega_{0} \in \cdot)$ has a
  bounded Lebesgue density $h$. Fix $E' < 0$ and $\lambda >0$. Then,
  for any $s\in]0,1[$, there is a constant $0\le
  Q_s=Q_s(\lambda,u,E')<\infty$ such that for all $\alpha \ge 0$, all
  energies $E \le E'$, all lengths $L>0$ and all $\eta\le |E'|/4$, one
  has
  \begin{equation}
    \label{wegner}
     \sup_{x \in\mathbb{Z}^{d}} \E \Bigl[ \tr \left(
       P_{\Lambda_L(x)}(J_\eta) \right)\Bigr]   \le  Q_s \eta^s L^d,
  \end{equation} 
  where $J_\eta :=[E-\eta,E+\eta]$ and $P_{\Lambda_L(x)}(J_{\eta}) :=
  \Chi_{J_{\eta}}\bigl(H_{\alpha,\lambda,\omega} (\Lambda_L(x))\bigr)$
  is the spectral projection of
  $H_{\alpha,\lambda,\omega}(\Lambda_L(x))$ associated with the
  interval $J_{\eta}$.  As a consequence,
  \begin{equation}
    \label{modifW2}
    \sup_{x \in\mathbb{Z}^{d}}
    \P \Bigl[ \dist\Bigl\{ \sigma\Bigl(H_{\alpha,\lambda,\omega}
    \bigl(\Lambda_L(x)\bigr)\Bigr) ,  E \Bigr\}
    \le \eta\Bigr] 
    \le
    Q_s \eta^s L^d.
  \end{equation}
\end{lemma}

\begin{remark}
  If the single site potential covers the unit cube,
  then \cite{CH} applies and one gets \eqref{wegner} with $s=1$ (see
  also \cite{CHK2} for a recent development). 
\end{remark}

\begin{remark}
  One might have expected a volume correction in \eqref{wegner} due to
  the geometry of the potential. That this is not the case relates to
  the fact that we consider only energies below the spectrum of
  $-\Delta$. The decaying envelope makes it even harder to get an
  eigenvalue close to a given $E \le E' < 0$. Actually, if the box
  $\Lambda_L(x)$ is far enough away from the origin, i.e.\ if
  \begin{equation} 
    \label{region} 
    \inf_{y \in\Lambda_{L}(x)} \scal{y}
    \ge \left(\frac{2\lambda u_0}{|E'|}\right)^{1/\alpha},
  \end{equation}
  then $H_{\alpha,\lambda,\omega}\bigl(\Lambda_L(x)\bigr)\ge \frac12
  E'$ and $\E \left[\tr \left( P_{\Lambda_L(x)}(J_\eta)
    \right)\right] = 0$.
\end{remark}

\begin{proof}[Proof of Lemma~\ref{lemwegner}.]
  We follow the strategy of \cite{CHN,CHK}. For convenience, let us
  write $d_0:=|E'|$ and $R_0(E) := (-\Delta - E)^{-1}$. Since
  $\dist(E,\sigma(-\Delta))\ge d_0$, one has
  \begin{align}
    \tr \{P_{\Lambda_L(x)}(J_\eta)\}   = & \tr \Bigl\{P_{\Lambda_L(x)}(J_\eta)
    \bigl(H_{\alpha,\lambda,\omega}(\Lambda_L(x))  -
    E\bigr) P_{\Lambda_L(x)}(J_\eta) R_0(E) \Bigr\}\notag \\  
     &    -\tr \bigl\{ P_{\Lambda_L(x)}(J_\eta) V_{\Lambda_L(x)}
    R_0(E) \bigr\} \notag\\
     \le & \frac{\eta}{d_0} \tr \{P_{\Lambda_L(x)}(J_\eta)\}  - \tr\bigl\{
    P_{\Lambda_L(x)}(J_\eta) V_{\Lambda_L(x)} R_0(E)\bigr\} . 
    \label{eqweg1} 
  \end{align}
  But notice that, using Cauchy-Schwarz and $\|R_0(E)\|\le  d_0^{-1}$,
  \begin{align}
    \bigl|\tr \bigl\{ P_{\Lambda_L(x)}(J_\eta) V_{\Lambda_L(x)} & 
      R_0(E)\bigr\} \bigr| \notag \\ 
    & \le   \frac1{d_0}\|P_{\Lambda_L(x)}(J_\eta)
    V_{\Lambda_L(x)}\|_2  \| P_{\Lambda_L(x)}(J_\eta)\|_2 \notag \\ 
    & \le  \frac1{2d_0^2} \tr \bigl\{P_{\Lambda_L(x)}(J_\eta)
    V_{\Lambda_L(x)}^2 \bigr\} 
    +  \frac12\tr\bigl\{ P_{\Lambda_L(x)}(J_\eta)\bigr\} .
    \label{eqweg2} 
  \end{align}
  Since we took $\eta\le d_0/4$, \eqref{eqweg1} and \eqref{eqweg2}
  combine to give
  \begin{equation}
    \tr \{P_{\Lambda_L(x)}(J_\eta) \}
    \le \frac{2}{d_0^2} \tr \bigl\{ P_{\Lambda_L(x)}(J_\eta)
    V_{\Lambda_L(x)}^2  \bigr\}
    \le \frac{2\lambda U_0}{d_0^2} \tr \bigl\{ P_{\Lambda_L(x)}(J_\eta)
    \tilde{V}_{\Lambda_L(x)}\bigr\}, 
    \label{eqweg3} 
  \end{equation}
  where $\tilde{V}_{\Lambda_L(x)} := \lambda \gamma_{\alpha} \sum_{i\in
    \Lambda_L(x) } u(\cdot - i)\ge 0$ and the constant $U_{0}$ was
  defined in part~(2) of Theorem~\ref{main}. The usual (but crucial)
  observation is that
  \begin{equation}
    \tilde{V}_{\Lambda_L(x)} = - \sum_{i\in \Lambda_L(x) }
    \frac{\partial V_{\omega,\Lambda_L(x)}}{\partial \omega_i}.  
  \end{equation}
  Next we pick a continuously differentiable, monotone decreasing
  function $f_\eta: \mathbb{R} \rightarrow [0,1]$ such that
  $f_\eta(\xi)=1$ for $\xi\le E-2\eta$ and $f_\eta(\xi)=0$ for $
  \xi\ge E+2\eta$. In particular, this function can be chosen such
  that $\Chi_{J_\eta} \le - C \eta f'_\eta$ holds with some constant
  $C$, which is independent of $\eta$.
  It follows that (recalling also $\tilde{V}_{\Lambda_L(x)}\ge 0$)
  \begin{align}
    \E \tr \bigl\{ & P_{\Lambda_L(x)}(J_\eta)
      \tilde{V}_{\Lambda_L(x)}\bigr\}  \notag \\ 
    &\le   C\eta \sum_{i\in \Lambda_L(x) } \E \tr \left\{
    f'_\eta\bigl(H_{\alpha,\lambda,\omega}(\Lambda_L(x)) \bigr) \;
    \frac{\partial V_{\omega,\Lambda_L(x)}}{\partial \omega_i}\right\} \\ 
    &\le  C\eta \sum_{i\in \Lambda_L(x) } \E_{\omega_i^\perp}
    \int_0^1\!\d\omega_i\, h(\omega_i) \,\frac{\partial }{\partial
      \omega_i} \tr \Bigl\{f_\eta\bigl(
    H_{\alpha,\lambda,\omega}(\Lambda_L(x))  \bigr)\Bigr\}  \\ 
    &\le  C\eta \, \|h\|_\infty \sum_{i\in \Lambda_L(x) }
    \E_{\omega_i^\perp} \tr \Bigl\{ f_\eta
    \bigl(H^{(\omega_i=1)}_{\alpha,\lambda,\omega}(\Lambda_L(x))
    \bigr) -  f_\eta \bigl(H^{(\omega_i=0)}_{\alpha,\lambda,\omega}(\Lambda_L(x))
    \bigr)\Bigr\}.\nonumber\\[-1ex]
    \label{eqweg4} 
  \end{align}
  The average $\E_{\omega_i^\perp}$ in the above inequalities is over
  all random variables except $\omega_{i}$.
  Now, using the spectral shift function, it follows from \cite{CHN}
  (or see \cite[Eq.~(A12) -- (A.14)]{CHK}) that, for any $s\in]0,1[$,
  \begin{equation}
    \tr \Bigl\{ f_\eta
    \bigl(H^{(\omega_i=1)}_{\alpha,\lambda,\omega}(\Lambda_L(x))
    \bigr) -  f_\eta \bigl(H^{(\omega_i=0)}_{\alpha,\lambda,\omega}(\Lambda_L(x))
    \bigr)\Bigr\}
    \le C^{\prime} (\lambda u_0)^{1-s} \eta^{s-1} ,
    \label{eqweg5}
  \end{equation}
  uniformly for all $\alpha \ge 0$.
  The bound \eqref{wegner} follows from \eqref{eqweg3},
  \eqref{eqweg4}, and  \eqref{eqweg5}.
\end{proof}

We are now ready to prove part~(3) of Theorem~\ref{main}.

\begin{proof}[Proof of part~(3) of Theorem~\ref{main}.] 
  Since boxes are independent at a distance, and we have the Wegner
  estimate of Lemma~\ref{lemwegner}, it suffices to prove an
  initial-scale estimate. Again, this has to be done uniformly in the
  location of the center of the box, because the model lacks
  translation invariance. Then the bootstrap multiscale analysis of
  \cite{GK1} applies. From now on, we fix $\lambda >0$.
  
  The most common method to prove the initial-scale estimate consists
  in emptying the spectrum of the finite-volume operator
  $H_{\alpha,\lambda,\omega}\bigl(\Lambda_{L_0}(x)\bigr)$ in an
  appropriately chosen interval
  $I_{\lambda}=[E_0(\lambda),E_1(\lambda)]$ at the bottom of the
  spectrum, e.g., \cite{CH,Kl1,Kl3,GK2,GK3}. There one can find proofs
  that in the ergodic situation $\alpha =0$ and for $L_0$ large
  enough, one gets
  $\sigma\bigl(H_{0,\lambda,\omega}(\Lambda_{L_0}(x))\bigr) \subset
  [E_1(\lambda) + m_0,+\infty[$ for some $m_{0}>0$ and all $\omega$ in
  some set of sufficiently large probability. Adding the envelope will
  only lift the spectrum up, and thus
  $\sigma\bigl(H_{\alpha,\lambda,\omega}(\Lambda_{L_0}(x))\bigr)\subset
  [E_1(\lambda) + m_0,+\infty[$ holds uniformly in $\alpha\ge 0$ and
  $x\in\R^d$ with sufficiently large probability (independently of
  $\alpha$). The Combes--Thomas estimate then provides the needed
  decay on the resolvent.

  Having the initial-scale estimate and the Wegner estimate at hand,
  the bootstrap multiscale analysis of \cite{GK1} can be performed.
  This provides for $\P$-a.e.\ $\omega$ the exponential decay of the
  eigenfunctions given in \eqref{sule}.  By a center of localization
  of an eigenfunction $\varphi_{n,\alpha,\lambda,\omega}$ with energy
  $E<0$, we mean a point $x_{n,\alpha,\lambda,\omega}\in\Z^d$ such
  that
  $\|\Chi_{x_{n,\alpha,\lambda,\omega}}\varphi_{n,\alpha,\lambda,\omega}\|
  = \sup_{x\in\Z^d} \|\Chi_x\varphi_{n,\alpha,\lambda,\omega}\|$. To
  determine the location of such centers of localization, we proceed
  as follows. Set $L_E := \max\{1, (2\lambda u_0 |E|^{-1})^{\frac
    1\alpha}\}$. Assume that $|x_{n,\alpha,\lambda,\omega}|\ge
  (N+1)L_E$, with $N\ge 1$, and consider the box
  $\Lambda_{NL_E}(x_{n,\alpha,\lambda,\omega})$. The spectrum of
  $H_{\alpha,\lambda,\omega}\bigl(\Lambda_{L_E}(x_{n,\alpha,\lambda,\omega})\bigr)$
  is separated from $E$ by a gap of size at least $|E|/2$. We estimate
  $\|\Chi_{x_{n,\alpha,\lambda,\omega}}\varphi_{n,\alpha,\lambda,\omega}\|$
  by the resolvent of
  $H_{\alpha,\lambda,\omega}\bigl(\Lambda_{L_E}(x_{n,\alpha,\lambda,\omega})\bigr)$.
  In the terminology of \cite{GK1}, this is called (EDI). We use the
  fact that $x_{n,\alpha,\lambda,\omega}$ maximizes $\|\Chi_x
  \varphi_{n,\alpha,\lambda,\omega}\|$ and that the finite volume
  resolvent decays exponentially as $\exp(- N L_E |E|)$, by a
  Combes--Thomas type argument. This leads to an inequality of the
  form $1\lesssim \exp(- N L_E |E|)$, and thus to a contradiction if
  $N$ is large enough. It remains to estimate $N$, and then we get
  that $|x_{n,\alpha,\lambda,\omega}|\le (N+1)L_E$.

  If $L_E=(2\lambda u_0 |E|^{-1})^{\frac 1\alpha}> 1$ (which is only
  possible if $|E|< 2\lambda u_0$ and $\alpha\le 1$), then $N L_E |E|=
  2\lambda u_0 N(2\lambda u_0 |E|^{-1})^{\frac 1\alpha - 1}$, which
  goes to infinity as $\alpha\downarrow 0$. It is thus enough to take
  $N=(2\lambda u_0)^{-1} C$ with a large enough universal constant
  $C$.  This implies that $|x_{n,\alpha,\lambda,\omega}|\lesssim
  (2\lambda u_0)^{\frac 1\alpha -1} |E|^{-\frac 1\alpha}$. If $L_E=1$,
  then $N L_E |E|=N |E|$, and we require $N=C |E|^{-1}$, with a large
  enough universal constant $C$. In this case
  $|x_{n,\alpha,\lambda,\omega}|\lesssim |E|^{-1}$.
\end{proof}

%%%%%%%%%%%%%%%%%%%%%%%%%%%%%%%%%%%%%%%%%%%%%%%%%%%%%%%%%%%%%%%%%%%%%%%%%%%%%%%
%%%%%%%%%%%%%%%%%%%%%%%%%%%%%%%%%%%%%%%%%%%%%%%%%%%%%%%%%%%%%%%%%%%%%%%%%%%%%%%
%%%%%%%%%%%%%%%%%%%%%%%%%%%%%%%%%%%%%%%%%%%%%%%%%%%%%%%%%%%%%%%%%%%%%%%%%%%%%%%

\begin{acknowledgement}
F.G.\ thanks the hospitality of
UC Irvine where this work was done. F.G.\ also thanks P.\ Hislop for
enjoyable discussions. 
\end{acknowledgement}

%%%%%%%%%%%%%%%%%%%%%%%%%%%%%%%%%%%%%%%%%%%%%%%%%%%%%%%%%%%%%%%%%%%%%%%%%%%%%%%
%%%%%%%%%%%%%%%%%%%%%%%%%%%%%%%%%%%%%%%%%%%%%%%%%%%%%%%%%%%%%%%%%%%%%%%%%%%%%%%
%%%%%%%%%%%%%%%%%%%%%%%%%%%%%%%%%%%%%%%%%%%%%%%%%%%%%%%%%%%%%%%%%%%%%%%%%%%%%%%

\end{document}